\documentclass[options]{JHEP3}

\title{Transgressions and Holographic Conformal Anomalies for Chern-Simons Gravities}
\author{Pablo Mora\\Instituto de F\'{\i}sica, Facultad de Ciencias, Igu\'a 4225,
Montevideo, Uruguay\\E-mail: pablmora-at-gmail-dot-com}

\abstract{ I present two calculations of the holographic Weyl anomalies   
induced by Chern-Simons gravity theories alternative to the ones presented in the literature. 
The calculations presented here rest on the extension from Chern-Simons to Transgression forms as lagrangians, motivated by gauge invariance, which automatically yields the boundary terms suitable to regularize the theory.
The procedure followed here sheds light in the structure of Chern-Simons gravities and their regularization.}

\begin{document}


\section{Introduction}

Chern-Simons (CS) gravities in 2+1 dimensions were introduced and studied in Ref.\cite{chern-simons-2+1}, extended to higher dimensions by Chamseddine in Refs.\cite{chamseddine} and to the supersymmetric case in Refs.\cite{banados-troncoso-zanelli}. These theories have been further studied and extended in several aspects aspects in many works, having very interesting properties from the point of view of their dynamic and symmetries, as they are true gauge theories of gravity with black hole, brane and other solutions. For a recent review of this topic with an extensive and comprehensive list of references see \cite{zanelli-lectures} (for older reviews see \cite{troncoso-zanelli}).

Chern-Simons forms are not strictly invariant under gauge transformations, but only quasi-invariant, meaning that they change by a closed form. Transgression forms (see for instance \cite{nakahara, alvarez}) are extensions of Chern-Simons forms that are strictly gauge invariant, but are functionals of two gauge fields $A$ and $\overline{A}$, unlike CS forms wich depend only on one gauge field $A$. Transgressions have been considered as actions for physical theories in refs.\cite{potsdam,transgression-branes,more-transgressions,IRS,sarda,tesis,motz1,motz3}, where several aspects of this models have been explored. In particular in refs.\cite{motz1,motz3} it was shown that the extensions of Chern-Simons gravities dictated by the transgressions have the built-in boundary terms that regularize the action, in the sense of giving a finite action, finite Noether conserved charges and the right black hole thermodynamics, unlike what happens in CS theories, where those quantities are infinite unless one regularizes them by hand.

In the context of the AdS-CFT correspondence \cite{maldacena,witten-ads,Gubser} gravitational theories in manifolds of dimension $d$ (bulk) which are asymptotically Anti de Sitter (AdS) and with Dirichlet boundary conditions are conjectured to be dual to certain Conformal Field Theories (CFT) in dimension $d-1$ (boundary). One of the first non trivial tests of this correspondence, suggested in \cite{witten-ads} and carried out in \cite{henningson} was to check if the Conformal or Weyl anomaly \cite{duff-weyl-anomaly} of the boundary theory matches the one induced by the bulk theory, as it should and indeed is. This check referred to General Relativity with cosmological constant regulated by suitable boundary terms in 5D in the bulk side and super Yang-Mill theory with $\mathcal{N}=4$ and a large number of colors (related to the cosmological constant of the bulk) in 4D in the boundary. After that conformal anomalies induced by higher curvature gravitational theories in 5D where computed following similar methods \cite{nojiri-odintsov,blau-narain-gava,schwimmer-theisen}.

The conformal anomaly induced by Chern-Simons gravitational theories was computed in \cite{banados-schwimmer-theisen} for 5D and 3D CS gravity (inducing Weyl anomalies in 4D and 2D CFTs respectively) and the generic form for arbitrary dimension was conjectured. This calculation was done only on the gravitational side, as the dual CFT theories are not known, but the result of doing this gravitational computation using several different methods was the same, and it was in agreement with what was to be expected from refs.\cite{nojiri-odintsov,blau-narain-gava,schwimmer-theisen}. The Weyl anomaly in 4D induced by CS gravity in 5D was also computed in \cite{banados-miskovic-theisen} while the Weyl anomaly induced by CS gravities in any dimension was computed in \cite{banados-olea-theisen}. In those works the Weyl anomaly was computed as the trace of the boundary energy-momentum tensor, adding counterterms to cancel infinite contributions.

In this paper I present two alternative computations of the Weyl anomaly for Chern-Simons gravities in any dimension from the point of view of transgressions form which make explicit the underlying geometrical structure, the role of the $\overline{A}$ (see below) as a regulator and provide further evidence for the claim of refs.\cite{motz1,motz3} that the proper boundary terms for CS gravities are the ones dictated by the gauge invariance (i.e. by going from CS to transgressions forms) which provide a built-in regularization. The explicit introduction of the second gauge field also allows to interpret the anomaly as coming from the non variance of the regulator.  

The first computation takes advantage of the gauge structure of CS gravity, and consist in taking a particular gauge transformation that would induce Weyl transformations in the boundary, varying only the first $A$ but not the non dynamical regulator $\overline{A}$. The anomaly can be read from the variation of the action under this transformation.

The second computation consist in taking the variation of the action under diffeomorphisms in the radial direction, which also induce Weyl transformations on the boundary\cite{imbimbo}, again reading the anomaly from this variation. Here one can 
explain the breaking of the symmetry by the presence of the boundary, but also observe that the field $\overline{A}$ has zero variation in this case. 

I consider that this calculations shed light from a different angle on the results of 
refs.\cite{banados-schwimmer-theisen,banados-olea-theisen,banados-miskovic-theisen}.

\section{Review of Transgressions in Field Theory}

This section is esentially a review of \cite{motz1,motz3}, see these references for more 
details and an exhaustive list of references.

\subsection{Transgressions}

Chern-Simons forms\footnote{For the details of the mathematics of Chern-Simons and Transgression forms see\cite{nakahara}.} $\mathcal{C}_{2n+1}(A)$ are differential forms defined for
a connection $A$, which under gauge transformations of that connection transform by a closed form, so are say to be \textit{quasi invariant}. Transgression forms $\mathcal{T}_{2n+1}$ are a generalization of Chern-Simons forms that are strictly invariant.  The use of this forms as lagrangians for physical theories, or as a template for actions for physical theories was discussed in references \cite{motz1, motz3}. Transgressions depend on two connections, $A$ and $\overline{A}$, and can be written as the difference of two Chern-Simons formas plus an exact form 
\begin{equation}
\mathcal{T}_{2n+1}=\mathcal{C}_{2n+1}(A)-\mathcal{C}_{2n+1}(\overline
{A})-dB_{2n}\left(  A,\overline{A}\right)  . \label{Transgression+CS+C2n}%
\end{equation}
or also as (see e.g., \cite{nakahara}),
\begin{equation}
\mathcal{T}_{2n+1}\left(  A,\overline{A}\right)  =(n+1)\int_{0}^{1}%
dt\ <\Delta{A}F_{t}^{n}>
\end{equation}
where\footnote{Here wedge product between forms is assumed.} 
$A_{t} = tA+(1-t)\overline{A}=\overline{A}+t\Delta A $
is a connection that interpolates between the two independent gauge potentials
$A$ and $\overline{A}$. The Lie algebra-valued one-forms\footnote{Notation: In what follows upper case latin indices from the beginning of the alphabet $A,~B,~C,...$ are space-time indices with values from $0$ to $d-1=2n$; upper case latin indices from the middle of the alphabet $I,~J,~K,...$ are space-time indices with values from $0$ to $d-1=2n$ but different from 1 (with 1 corresponding to a "radial" coordinate, or a coordinate along the direction normal to the boundary); lower case latin indices from the beginning of the alphabet $a,~b,~c,...$ are tangent space (or Lorentz) indices with values from $0$ to $d-1=2n$; lower case latin indices from the middle of the alphabet $i,~j,~k,...$ are tangent space (or Lorentz) indices with values from $0$ to $d-1=2n$ but different from 1 (with 1 identified to a "radial" direction, or a direction normal to the boundary in tangent space). The index $\alpha$ labels the generators $G_{\alpha}$ of the Lie group considered and takes values from 1 to the dimension of the group.} $A=A_{A}^{\alpha}G_{\alpha}\ dx^{A}$ and $\overline{A}=\overline{A}_{A}^{\alpha}G_{\alpha}\ dx^{A}$ are
connections under gauge transformations, $G_{\alpha}$ are the generators, and
$<\cdots>$ stands for a symmetrized invariant trace in the Lie algebra. 
The corresponding curvature is
$F_{t}=dA_{t}+A_{t}^{2}=t F +(1-t)\overline{F}-t(1-t)(\Delta A)^{2}$.  Setting $\overline{A} =0$ 
in the transgression form yields the Chern-Simons form for $A$.

\subsection{Chern-Simons and Transgression Gravity}

For the AdS group in dimension $d=2n+1$ the gauge connection is given by\footnote{A gauge
connection has dimensions of $(lenght)^{-1}$, so it must be  
$A=\frac{\omega ^{ab}}{2}J_{ab}+\frac{e^a}{l}P_a$ where $l$ is the 'AdS radius'.
I set $l=1$ trough all the present paper. It is
easy to reintroduce $l$ using dimensional
analysis, if necessary.}
$A=\frac{\omega ^{ab}}{2}J_{ab}+e^aP_a$ where $\omega ^{ab}$ is the spin connection, $e^a$ is the vielbein and $J_{ab}$ and $P_a$
are the generators of the AdS group (for Lorentz transformations and
translations respectively).  One possible symmetrized trace, and the only one I will consider in this paper, is $$<J_{a_{1}a_{2}}...J_{a_{2n-1}a_{2n}}P_{a_{2n+1}}>=\kappa\frac{2^{n}}
{(n+1)}\epsilon_{a_{1}...a_{2n+1}}$$

Then the transgression for the AdS group  is\footnote{In what follows I will use a
compact notation where $\epsilon$ stands for the Levi-Civita symbol
$\epsilon _{a_1...a_d}$ and wedge products of differential forms are
understood, as it was done in Refs.\cite{motz2,motz1,motz3}. For instance: $\epsilon Re^{d-2}\equiv \epsilon
_{a_1a_2....a_d}R^{a_1a_2}\wedge e^{a_3}\wedge ...\wedge
e^{a_{d-2}}$,  $(\theta ^2)^{ab}=\theta ^a_c\wedge\theta ^{cb}$.} \cite{motz3}
\begin{equation}
{\cal T}_{2n+1}= \kappa \int _0^1dt \epsilon (R+t^2e^2)^ne -\kappa
\int _0^1dt \epsilon (\tilde{R}+t^2\overline{e}^2)^n\overline{e}
+d~B _{2n}
\end{equation}
where
\begin{equation}
B _{2n}=-\kappa n\int_0^1dt\int_0^1ds~\epsilon\theta e_t
\left\{t R+(1-t)\tilde{R}-t(1-t)\theta ^2+s^2e_t^2 \right\}^{n-1}
\end{equation}
Here $e ^a$ and $\overline{e}^a$ are the two vielbeins and $\omega
^{ab} $ and $\overline{\omega}^{ab}$ the two spin connections,
$R=d\omega +\omega ^2$ and
$\tilde{R}=d\overline{\omega}+\overline{\omega}^2$ are the
corresponding curvatures, $\theta =\omega -\overline{\omega}$ and
$e_t=te+(1-t)\overline{e}$. Written in a more compact way
\begin{equation}
B _{2n}=-\kappa n\int_0^1dt\int_0^1ds~\epsilon\theta e_t
\overline{R} _{st}^{n-1}
\end{equation}
where
$$\overline{R} _{st}=t R+(1-t)\tilde{R}-t(1-t)\theta ^2+s^2e_t^2$$

The action for transgressions for the AdS group is chosen to be
\cite{motz3}
\begin{equation}
I_{Trans}= \kappa \int _{\cal M}\int _0^1dt \epsilon (R+t^2e^2)^ne
-\kappa\int_{\overline{{\cal M}}} \int _0^1dt \epsilon
(\tilde{R}+t^2\overline{e}^2)^n\overline{e} +\int _{\partial {\cal
M} }B _{2n}
\end{equation}
where ${\cal M}$ and $\overline{\cal M}$ are two manifolds with a
common boundary, that is $\partial {\cal M}\equiv\partial
\overline{{\cal M}}$. Notice that this is a generalization from he
simpler case where ${\cal M}\equiv\overline{\cal M}$.  

\subsection{The background independent vacuum}

The natural, or rather the naive, vacuum in a field theory corresponds to the configuration in which all fields vanish. One may regard the transgression as a tool to regularize a physical theory, with $A$ being the physical fields and $\overline{A}$ some non dynamical regulator configuration or vacuum to be substracted. The choice $\overline{A}=0$ (the naive vacuum), which yields Chern-Simons forms, gives infinite values for conserved charges and thermodynamic quantities for black holes as well as an ill-defined action principle. In refs.\cite{motz1,motz3} it was shown that a choice of $\overline{A}$ that properly regularize the action is given by
\begin{equation}
\overline{\omega}^{ij}=\omega ^{ij},\;\;\;\overline{\omega}^{1j}=0 ,\;\;\; \overline{e}=0
\end{equation}
Notice that the vielbein does vanish for this configuration, but the spin connection does not. 
We have then $\Delta A= \frac{1}{2}\theta J+eP$ and $\Delta A^2=\frac{1}{2}(\theta ^2+e^2)J+\theta eP$. 
For this choice of $\overline{A}$ we have that $\overline{T}^a=0$ and the identities 
$$R^{ij}=\overline{R}^{ij}+(\theta ^2)^{ij}~~~,~~~
R^{1i}=d\theta ^{1i}+\overline{\omega}^i_k\theta ^{1k}=\overline{D}\theta ^{1i}$$ 
which are just the Gauss-Codazzi equations. For this $\overline{A}$ configuration the boundary term reduces to
$$
\mathcal{B}_{2n}=-\kappa n\int_{0}^{1}dt\int_{0}^{t}ds~\epsilon~\theta\ e\left(  \bar
{R}+t^{2}\theta^{2}+s^{2}e^{2}\right)  ^{n-1}\ ,
$$
and the action becomes
\begin{equation}
\mathcal{I}_{Trans}= \kappa \int _{\cal M}\int _0^1dt \epsilon (R+t^2e^2)^ne
+\int _{\partial {\cal M} }\mathcal{B} _{2n}
\end{equation}
This action is the one we are interested in, consisting of a bulk 
Chern-Simons gravity term plus a boundary term that regularizes 
and makes well defined the action, coming from its origin as a transgression form.
 The action of eq.(2.6) and the action of eq.(2.8) yield exactly the same equations 
 of motion for $A$, but the $\overline{A}$ chosen in this subsection does not 
 satisfy the equations of motion corresponding to the action of eq.(2.6) unless the 
 Euler density of the boundary vanish, and must be regarded as a non dynamical field in 
 the context of eq.(2.8). The proper boundary conditions for the action of eq.(2.8) 
 where discussed in \cite{motz1,motz3}.

Quite remarkably, the boundary terms $\mathcal{B} _{2n}$, with a suitable constant coefficient,
also works as a regulator counterterm for AdS gravity (General Relativity with cosmological constant) in odd dimensions \cite{motz2}, and in 
fact for any Lovelock AdS gravity \cite{olea-kounterterms,kofinas-olea}. This is a rather 
surprising result, as one would expect a counterterm boundary term to be a polynomial in the intrinsic and extrinsic curvatures of the boundary with relative coefficients differing from case to case, instead of having always the same polynomial in a given dimension, changing only its global coefficient. 

The boundary term $B_{2n}$ of the previous subsection, with a suitable constant coefficient, is a suitable regulator for odd-dimensional AdS gravity in a background substraction approach , as shown in ref.\cite{transgression-gr}, where it was argued that the AdS gravity action with that boundary term an 
coefficient has enhaced symmetry properties (an analogous weaker enhancement of symmetry than the one achieved in passing from CS to transgression forms). 

\section{Chern-Simons Holographic Conformal Anomaly}

\subsection{Gauge variation calculation}

The variation of the transgression is
\begin{equation}
\delta\mathcal{T}_{2n+1}=(n+1) <F^n\delta A>-<\overline{F}^n\delta \overline{A}>
-n(n+1)d\{\int _0^1dt<\Delta AF_t^{n-1}\delta A_t>\}
\end{equation}
Under gauge transformations that change $A$ but not $\overline{A}$ so that $\delta A=D\lambda$ and $\delta \overline{A}=0$ we have
\begin{equation}
\delta _{\lambda}\mathcal{T}_{2n+1}=+(n+1) <F^nD\lambda> 
-n(n+1)d\{\int _0^1dt t<\Delta AF_t^{n-1}D\lambda>\}
\end{equation}
The first term of the second member vanish as a consequence of the equations of motion of the theory $<F^nG>=0$ where $G$ is a generator of the group (see for instance \cite{motz3}), while integration by parts in the second member yields
\begin{equation}
\delta _{\lambda}\mathcal{T}_{2n+1}=-n(n+1)d\{\int _0^1dt t<\left[D(\Delta A)F_t^{n-1}-(n-1)\Delta A(DF_t)F_t^{n-2}\right]\lambda>\}
\end{equation} 
Using $D(\Delta A)=F-\overline{F}+\Delta A^2$ and $DF_t=(1-t)[\Delta A,F_t]$ we get
\begin{equation}
\delta _{\lambda}\mathcal{T}_{2n+1}=-n(n+1)d\{\int _0^1dt t<\left[(F-\overline{F}+\Delta A^2) F_t^{n-1}
-(n-1)(1-t)\Delta A[\Delta A,F_t] F_t^{n-2}\right]\lambda>\}
\end{equation} 
Given the vielbeins $e$, $\overline{e}$ and the spin connections $\omega$, $\overline{\omega}$, we have $F=\frac{1}{2}(R+e^2)J+TP$ and $\overline{F}=\frac{1}{2}(\overline{R}+\overline{e}^2)J+\overline{T}P$,
$\Delta A=\frac{1}{2}(\omega-\overline{\omega}) J+(e-\overline{e})P\equiv\frac{1}{2}\theta J+EP$. There the curvature is $R=d\omega +\omega ^2$and the torsion is $T=De=de+\omega e$ and similarly for the corresponding overlined objects.
For the choice of $\overline{A}$ of Subsection 2.3  
$$
\overline{\omega}^{ij}=\omega ^{ij},\;\;\;\overline{\omega}^{1j}=0 ,\;\;\; \overline{e}=0
$$
We therefore have, referring to Appendix A, $\overline{\omega}^{ij}=\hat{\omega}^{ij}$ and $\overline{R}^{ij}=\hat{R}^{ij}$.
Furthermore
$F-\overline{F}=\frac{1}{2}(\theta ^2+e^2)^{ij}J_{ij}+ \overline{D}\theta ^{1i}J_{1i}+T^aP_a$. We will need just 
the components along $J_{ij}$ of the following combinations
$$
(F-\overline{F}+\Delta A^2)^{ij}=(\theta^2+e^2)^{ij}
$$
$$F_t^{ij}=[tF+(1-t)\overline{F}-t(1-t)\Delta A^2]^{ij}=\frac{1}{2}[t(\overline{R}+\theta ^2+e^2)+(1-t)\overline{R}-t(1-t)(\theta^2+e^2)]^{ij} $$
or 
$$
F_t^{ij}=\frac{1}{2}[\overline{R}+t^2(\theta^2+e^2)]^{ij} 
$$ 

 The gauge transformations that generate conformal transformations at the boundary are of the form $\lambda =\sigma P_1$ where $\sigma$ is a function of the coordinates that is non-singular at the boundary (see Appendix B), as one may naively expect. The symmetrized trace is schematically $<J...JP>=\kappa \frac{2^n}{n+1}\epsilon$ and zero otherwise, that is the trace is non-vanishing if it involves $n$ $J$'s an one $P$ and zero if it involves only $J$'s or any number of $P$'s other than one. From the las expression for the gauge variation of the transgression one sees that if $\lambda =\sigma P_1$ then in the product inside the symmetrized trace only the components of all the other factors along $J$ and for indices other than 1 will matter. Notice in particular that the components of $\Delta A$ along $J$ for indices other than 1 is zero, implying that the second term inside the trace in $\delta _{\lambda}\mathcal{T}_{2n+1}$ vanish. 
For gauge transformations of this special form we get
\begin{equation}
\delta _{\lambda}\mathcal{T}_{2n+1}= -\kappa \frac{2^n}{n+1}d\{\frac{n(n+1)}{2^{n-1}}\int_0^1dt t\epsilon(\theta^2+e^2)[\overline{R}+t^2(\theta^2+e^2)]^{n-1}\sigma\}
\end{equation}
or, changing variables from $t$ to $u=t^2$
\begin{equation}
\delta _{\lambda}\mathcal{T}_{2n+1}= -\kappa \frac{2^n}{n+1}d\{\frac{n(n+1)}{2^{n}}\int_0^1du\epsilon(\theta^2+e^2)[\overline{R}+u(\theta^2+e^2)]^{n-1}\sigma\}
\end{equation}
Expanding the integrand and integrating term by term in $u$ it results
\begin{eqnarray}
n\int_0^1du\epsilon(\theta^2+e^2)[\overline{R}+u(\theta^2+e^2)]^{n-1}=\nonumber\\
=n\epsilon(\theta^2+e^2)\sum _{k=0}^{n-1}\frac{(n-1)!}{k!(n-1-k)!}\overline{R}^k(\theta^2+e^2)^{n-1-k}\left(\frac{u^{n-k}}{n-k}\right)\mid _0^1
\end{eqnarray}
or
\begin{eqnarray}
n\int_0^1du\epsilon(\theta^2+e^2)[\overline{R}+u(\theta^2+e^2)]^{n-1} 
=\epsilon\sum _{k=0}^{n-1}\frac{n!}{k!(n-k)!}\overline{R}^k(\theta^2+e^2)^{n-k}
\end{eqnarray}
This can be simplified using the equations of motion $<F^nG>=0$ in the case the generator $G$ is one of the translation generators $P$, which are
\begin{equation}
\epsilon(R+e^2)^n=0
\end{equation}
or
\begin{equation}
0=\epsilon(\overline{R}+(\theta ^2+e^2))^n=
\epsilon\sum _{k=0}^{n}\frac{n!}{k!(n-k)!}\overline{R}^k(\theta^2+e^2)^{n-k}=
\epsilon\overline{R}^n+\epsilon\sum _{k=0}^{n-1}\frac{n!}{k!(n-k)!}\overline{R}^k(\theta^2+e^2)^{n-k}
\end{equation}
Therefore
\begin{eqnarray}
n\int_0^1du\epsilon(\theta^2+e^2)[\overline{R}+u(\theta^2+e^2)]^{n-1} 
=-\epsilon\overline{R}^n
\end{eqnarray}
and we obtain
\begin{equation}
\delta _{\lambda}\mathcal{T}_{2n+1}= d\{\kappa\epsilon\overline{R}^{n}\sigma\}
\end{equation}
then the variation of the action of eq.(2.8) under gauge transformations of the kind considered, 
which induce Weyl transformations in the boundary, is
\begin{equation}
\delta\mathcal{I}_{Trans}= \int _{\partial {\cal M} } \kappa\epsilon\overline{R}^{n}\sigma
\end{equation}
But the anomaly $\mathcal{A}$ correspond to the non invariance of the action under whatever transformation 
one is considering, with the variation of the action of the generic form
\begin{equation}
\delta\mathcal{I}_{Trans}= \int _{\partial {\cal M} } \mathcal{A} \sigma
\end{equation}
therefore the Weyl anomaly is
\begin{equation}
\mathcal{A}= \kappa\overline{E}_n
\end{equation}
where $\overline{E}_n=\epsilon\overline{R}^{n}$ is the Euler density of the boundary.

\subsection{Radial diffeomorphisms calculation}

The strategy in this subsection is to compute the variation of our action eq.(2.8) under radial diffemorphisms
generated by the vector $\xi =\sigma \frac{\partial ~}{\partial r}$, which according to ref.\cite{imbimbo} generate Weyl transformations in the boundary. The generic variation should again be of the form
\begin{equation}
\delta\mathcal{I}_{Trans}= \int _{\partial {\cal M} } \mathcal{A} \sigma
\end{equation}
where $\mathcal{A}$ is the Weyl anomaly.

We need again the variation of the transgression
$$
\delta\mathcal{T}_{2n+1}=(n+1) <F^n\delta A>-<\overline{F}^n\delta \overline{A}>
-n(n+1)d\{\int _0^1dt<\Delta AF_t^{n-1}\delta A_t>\}
$$
Under infinitesimal diffeomosphisms generated by the vector 
$\xi =\xi ^{A}\frac{\partial ~}{\partial x^{A}}$ the variation of the gauge potentials is given by $\delta _{\xi}A=\mathcal{L}_{\xi}A$ and
$\delta _{\xi}\overline{A}=\mathcal{L}_{\xi}\overline{A}$. For gauge potentials this can be written as
$\delta _{\xi}A=D[I_{\xi}A]+I_{\xi}F$ and $\delta _{\xi}\overline{A}=\overline{D}[I_{\xi}\overline{A}]+I_{\xi}\overline{F}$, where $I_{\xi}$ is the contraction operator\footnote{The contraction operator acting on a p-form $\alpha _p$ is given by
$
I_{\xi}\alpha_{p}=\frac{1}{(p-1)!}\xi^{\nu}\alpha_{\nu\mu_{1}...\mu_{p-1}
}dx^{\mu_{1}}...dx^{\mu_{p-1}}
$
The operator $I_{\xi}$ is an antiderivation, in the sense that acting on the
exterior product of two differential forms $\alpha_{p}$ and $\beta_{q}$ of
orders $p$ and $q$ it gives $I_{\xi}(\alpha_{p}\beta_{q})=(I_{\xi}\alpha
_{p})\beta_{q}+(-1)^{p}\alpha_{p}(I_{\xi}\beta_{q})$}. We will use the same choice of $\overline{A}$ as in the previous section.
We will chose a gauge in which $\omega ^{ab}_r=0$ and $e^a_r=\delta ^a_r$, which is possible as this is the same number of conditions as gauge parameters in the {\cal AdS} group, and corresponds to the form explicitly given in Appendix A. Furthermore we are interested in radial diffeos, for which $\xi =\sigma \frac{\partial ~}{\partial r}$, therefore $I_{\xi}\omega=I_{\xi}\overline{\omega}=I_{\xi}\theta =0$ and $I_{\xi}e^a=\sigma \delta ^a_1$. Then $I_{\xi}A=\sigma P_1$ and $I_{\xi} \overline{A}=0$. Because $\overline{e}=0$ then $\overline{T}=0$ hence we have $I_{\xi}\overline{F}=\frac{1}{2}I_{\xi}\overline{R}J=0$, because $\xi$ is a radial vector and  $\overline{\omega}$ has no component along $dr$ nor dependence on $r$. It follows $\delta _{\xi}\overline{A}=0$. Then, using that the equation of motion for $A$ implies $<F^n\delta _{\xi}A>=0$ we get
\begin{equation}
\delta _{\xi}\mathcal{T}_{2n+1}= -d\{n(n+1)\int _0^1dt~t<\Delta AF_t^{n-1}D[I_{\xi}A]>+n(n+1)\int _0^1dt~t<\Delta AF_t^{n-1}I_{\xi}F>\}
\end{equation}
The second term of the second member of the previous equation is zero, because $I_{\xi}F=0$ for $\xi =\sigma\frac{\partial ~}{\partial r}$, as shown in Appendix A, while the first term of the second member, taking in account that $I_{\xi}A=\sigma P_1$, yields exactly the same expression obtained in the previous subsection, which following the same steps leads to the same expression for the conformal anomaly, as it should.

\section{Discussion and Comments}

It was shown that using the already regularized action eq.(2.8) and the fact that it is secretly a transgression
simple derivations of the Weyl anomaly are possible, which help understan its origin.
In subsection 3.1 the non variation of $\overline{A}$ results in the non invariance of the action and the anomaly, in agreement with the well known fact that anomalies arise from a non invariant regulator. On the other hand if one considers a dynamical $\overline{A}$ as in the action of eq.(2.6), which is varied under gauge 
transformations that agree for $A$ and $\overline{A}$ in the boundary (which would not preserve eq.(2.7)) the action would be invariant and the anomaly is gone. This is analogous to the Wess-Zumino mechanism(see for instance\cite{alvarez}) where the anomaly is cancelled at the cost of introducing new dynamical degrees of freedom.
In subsection 3.2 the variation of $\overline{A}$ is also zero, but one could more properly understand the anomaly
as due to the presence of the boundary.

Other manifestations of the Weyl anomaly that I have not discussed but are of course related (besides the non zero trace of the boundary energy-momentum tensor) are a non zero vacuum energy (computed in \cite{motz1}) or the appearance of central charges in the algebra of conserved charges, which could be obtained modifying the calculation of \cite{motz3} to the case where only one of the gauge fields transform.

The action considered here could be used to compute holographic currents and their eventual anomalies, wich would not require further regularization.

As mentioned at the end of section 2 it turns out that the boundary term borrowed from our transgression inspired action is the right one for Lovelock gravities. It would be interesting to do a near boundary analysis a la Fefferman-Graham for Lovelock theories, verify if there is an enhancement of symmetry that somehow explain that such boundary term works, and then carry out a calculation of the holographic currents and their possible anomalies
for those theories, along the lines of \cite{banados-miskovic-theisen}.

\acknowledgments{I am grateful to O. Miskovic and R. Olea for enlightening discussions and comments.
I acknowledge funding from the \textit{Sistema Nacional de Investigadores (SNI)}, Uruguay.} \newline

\centerline{\textbf{APPENDICES}}

\appendix

\section{Fefferman-Graham coordinates for Chern-Simons gravity}

This appendix reviews results described esentially in ref.\cite{banados-miskovic-theisen}, see this paper for further details and references.
The Fefferman-Graham \cite{fefferman-graham} form of the asymptotic metric for asymptotically AdS spaces (AAdS) is
\begin{equation}
ds^2=\frac{d\rho ^2}{4\rho ^2}+\frac{1}{\rho}\hat{g}_{IJ}dx^Idx^J
\end{equation}
with $\rho $ a radial coordinate such that $\rho =0$ corresponds to the boundary, $x^I$ with $I\neq 1$ are the "transverse coordinates", and the transverse 
metric $\hat{g}_{IJ}(x, \rho)$ is analytic in $\rho$
$$\hat{g}_{IJ}(x,\rho)=\sum_{n=0}^{\infty}\hat{g}^{(n)}_{IJ}(x)\rho ^n$$
The AdS metric corresponds to $\hat{g}_{IJ}=\hat{\eta}_{IJ}$. Another form of this metric, where the boundary correspond to infinite radius $r$ results if we define 
$r=-\frac{1}{2}\log \rho$, then 
\begin{equation}
ds^2=dr^2+e^{2r}g_{IJ}dx^Idx^J
\end{equation}
For a generic gravity theory in AAdS spacetimes with Dirichlet boundary conditions in the metric one can determine the coefficients of the higher order terms $\hat{g}^{(n)}_{IJ}$ with $n\geq 1$ in terms of the boundary metric $\hat{g}^{(0)}_{IJ}$.

If $g_{IJ}=\hat{\eta } _{ij}\hat{e}^i_I\hat{e}^j_J$
then the vielbein one-form is 
$$e^1=dr~~~,~~~e^i=e^{r}\hat{e}^i=\frac{1}{\sqrt{\rho}}\hat{e}^i$$
where $\hat{e}^i=\hat{e}^i_Idx^I$.
For a generic first order gravity theory in an AAdS spacetime the vielbein would then be of the form
$$e^i(x,\rho )=\frac{1}{\sqrt{\rho}}\sum_{n=0}^{\infty}\hat{e}^{(n)i}(x)\rho ^n$$
and similarly the spin connection would be
$$\omega ^{ij}(x,\rho )=\frac{1}{\sqrt{\rho}}\sum_{n=0}^{\infty}\hat{\omega}^{(n)ij}(x)\rho ^n$$
Remarkably, for the kind of Chern-Simons gravities discussed above this expressions were shown to be truncated, with only a finite number of coefficients of the expansion being non zero. The explicit expressions are
\begin{equation}
e^1=dr~~~,~~~\omega^{ij}(x,\rho )=\hat{\omega}^{ij}(x)
\end{equation}
\begin{equation}
e^i(x,\rho )=\frac{1}{\sqrt{\rho}}(\hat{e}^{(0)i}(x)+\rho \hat{e}^{(1)i}(x) )~~~,~~~\omega^{i1}(x,\rho )=\frac{1}{\sqrt{\rho}}(\hat{e}^{(0)i}(x)-\rho \hat{e}^{(1)i}(x) )
\end{equation}
We have for the components of the torsion $T^a=de^a+\omega ^a_be^b$
$$T^1=-2\hat{e}^{(0)}_i\hat{e}^{(1)i}~~~,~~~T^i=\frac{1}{\sqrt{\rho}}(\hat{T}^i+\rho \hat{D}\hat{e}^{(1)i})$$
where 
$\hat{T}^i=d\hat{e}^{(0)i}+\hat{\omega} ^i_j\hat{e}^{(0)j}$ is the boundary torsion and $\hat{D}\hat{e}^{(1)i}=d\hat{e}^{(1)i}+\hat{\omega ^i_j}\hat{e}^{(1)j}$. 
The components of the curvature $R^{ab}=d\omega ^{ab}+\omega ^a_c\omega ^{cb}$ are
$$R^{ij}=\hat{R}^{ij}-e^ie^j+2(\hat{e}^{(0)i}\hat{e}^{(1)j}-\hat{e}^{(0)j}\hat{e}^{(1)i})~~~,
~~~R^{1i}=-e^1e^i+\frac{1}{\sqrt{\rho}}(\hat{T}^i-\rho \hat{D}\hat{e}^{(1)i})$$
where $\hat{R}^{ij}=d\hat{\omega} ^{ij}+\hat{\omega} ^i_k\hat{\omega ^{kj}}$ is the boundary Riemann curvature. 
The components of the field strength $F=\frac{1}{2}(R^{ab}+e^ae^b)J_{ab}+T^aP_a$
are $F^i=T^i$, $F^1=T^1$, with the torsions given above, and 
$$F^{ij}=\hat{R}^{ij}+2(\hat{e}^{(0)i}\hat{e}^{(1)j}-\hat{e}^{(0)j}\hat{e}^{(1)i})~~~,
~~~F^{1i}=\frac{1}{\sqrt{\rho}}(\hat{T}^i-\rho \hat{D}\hat{e}^{(1)i})$$
For use in the main text it is important to notice that the matrix valued two-form $F$ 
has no component along $dr$, hence $I_{\xi}F=0$ for $\xi =\sigma\frac{\partial ~}{\partial r}$.

\section{Gauge transformations and Weyl transformations}

The equations A.3 and A.4 can be written as
\begin{equation}
A=\frac{1}{2}\omega^{ij}J_{ij}+e^{1}P_{1}+\frac{1}{\sqrt{\rho}}\hat{e}^{(0)i}J^{+}_i+\sqrt{\rho}\hat{e}^{(1)i}J^{-}_i
\end{equation}
where $J^{\pm}_i=P_i\pm J_{1i}$. Under gauge transformations $\delta A=D\lambda=d\lambda +[A,\lambda ]$. For $\lambda =\sigma P_1$, using the algebra of the AdS group generators we get $A'=A+\delta A$ of the form of eq. B.1, with
\begin{equation}
\delta e^{1}=d\sigma~~~,~~~\delta\hat{e}^{(0)i}=\sigma\hat{e}^{(0)i}
\end{equation}
\begin{equation}
\delta\omega ^{ij}=0~~~,~~~\delta\hat{e}^{(1)i}=-\sigma\hat{e}^{(1)i}
\end{equation}
Considering that $\hat{e}^{(0)i}$ is the vielbein of the boundary, the second equation of A.2 shows that gauge transformations with $\lambda =\sigma P_1$ do indeed generate Weyl transformations in the boundary.

\end{document}